\documentclass[twocolumn]{aa}
\def\l{{\ell}}
\def\lm{\l m}
\def\alm{a_{\l m}}
\def\cl{C_{\l}}
\def\dl{D_{\l}}
\def\bl{B_{\l}}

\begin{document}

\title{First Release of Gauss--Legendre Sky Pixelization (GLESP)
  software package for CMB analysis}

\author{
A.G. Doroshkevich$^{1}$, P.D. Naselsky$^{2,3}$,
O.V. Verkhodanov$^{4}$, D.I. Novikov$^{5,1}$, V.I. Turchaninov$^{6}$, \\
I.D. Novikov$^{1,2}$, P.R. Christensen$^{2}$, L.-Y. Chiang$^{2}$
}

\institute{ 
     Astro Space Center of Lebedev Physical Institute, Profsoyuznaya 84/32,
      Moscow, Russia 
\and Niels Bohr Institute, Blegdamsvej 17, DK-2100 Copenhagen, Denmark   
\and Rostov State University, Zorge 5, Rostov-Don, 344090, Russia
\and Special Astrophysical Observatory, Nizhnij Arkhyz, Karachaj-Cherkesia,
      369167, Russia   
\and Astrophysics, the Blackett Laboratory, Imperial College, London SW7 2AZ, UK
\and Keldysh Institute of Applied Math, Russian Academy of Science,
      125047 Moscow, Russia
}

\date{}

\abstract{
We report the release of the Gauss--Legendre Sky Pixelization
(GLESP) software package version 1.0. In this report we present the
main features and functions for processing and manipulation of sky
signals. Features for CMB polarization is underway and to be
incorporated in a future release. Interested readers can visit {\tt
  http://www.glesp.nbi.dk/} and register for receiving the package. 

}
\maketitle

\markboth
{Doroshkevich et~al.: First Release of GLESP software package}
{Doroshkevich et~al.: First Release of GLESP software package}

\section{Introduction}

The CMB temperature fluctuations on a sphere can be written as a sum over spherical harmonics
\begin{equation}
\Delta T(\theta,\varphi)=\sum^{\infty}_{\l=2} \sum^{\l}_{m=-\l} |\alm|e^{i\phi_{\lm}} Y_{\lm}
(\theta,\varphi), 
\label{eq1}
\end{equation}
where $|\alm|$ are the moduli of the spherical harmonic 
coefficients of the expansion  and $\phi_{\lm}$ are the phases defined in $[0, 2\pi]$. The angular power spectrum is defined as
\begin{equation}
\cl = \frac{1}{2 \l+1} \sum^{\l}_{m=-\l} |\alm|^2
\end{equation}
and 
\begin{equation}
\dl = \frac{\l(\l+1)}{2 \pi}\frac{\cl}{T_0}.
\end{equation}
where $T_0$ is the CMB temperature 2.726 K.

In this article, we report the first release of the Gauss--Legendre
Sky Pixelization (GLESP) software package for the CMB analysis. The GLESP
scheme (\cite{glesp}) is based on the Gauss--Legendre polynomial zeros,
allowing one to perform strict orthogonal spherical harmonic expansion
of the CMB temperature anisotropies. The GLESP package has been tested
on the following operating systems: RedHat Linux, SGI Irix and
Solaris. Features for CMB polarization are under preparation.

\section{Features in the GLESP package v1.0}
\subsection{Basic features for CMB analysis}
The basic features of the GLESP package includes spherical harmonic
analysis of a map, map synthesis from a given set of $\alm$
coefficients, and simulation of a Gaussian map from a given angular
power spectrum $\cl$, or $\dl$. 

\begin{description}
\item{\bf cl2map}

{\tt cl2map} synthesizes CMB temperature maps from spherical harmonic coefficients $\alm$. Inversely, it can decompose an input CMB temperature map into its spherical harmonic coefficients $\alm$ and/or power spectrum. 
If the input is a power spectrum $\cl$, or $\dl$, {\tt cl2map} generates a Gaussian map of CMB anisotropies by assigning the $\alm$ mutually independent Gaussian-distributed real and imaginary parts, both with a standard deviation $ \sqrt{\cl/2}$ at each $\l$ harmonic.

\item{\bf rsalm}

{\tt rsalm} convolves or deconvolves a map with a symmetric beam via
\begin{equation}
a^{'}_{\lm} = \bl \, \alm,
\end{equation}
where $\bl$ is the expansion in Legendre polynomials of a symmetric beam.
The symmetric beam can be input by reading in an ASCII file
containing the profile of $\bl$ or by providing the FWHM for
a symmetric Gaussian beam, where $\bl = \exp[- \sigma^2 \l (\l+1) / 2 ]$ and $\sigma = {\rm FWHM} / \sqrt{8 \ln 2}$. 

For deconvolution, a simple Tikhonov regularization scheme is applied via
\begin{equation}
a^{''}_{\lm} = \frac{\alm}{\bl + \alpha},
\end{equation}
where $\alpha$ is the regularization parameter. 

\item{\bf f2fig}

{\tt f2fig} converts a GLESP map in FITS binary format into GIF (Graphic Interchange Format).  
{\tt f2fig} offers users to put markers on the output figure. The
markers can be input with one of the following 3 coordinate systems:
in $\theta$ and $\phi$ of spherical polar coordinate system, in Galactic latitude $b$ and longitude $l$ of Galactic coordinate system and in right ascension (R.A.) and declination (Dec.) of equatorial coordinate system.

\item{\bf alm2dl}

{\tt alm2dl} reads $\alm$ and calculates the angular power spectrum $\cl$ or $\dl$. It also provides the options in listing the amplitudes $|\alm|$ or phases $\phi_{\lm}$ defined in $[0, 2\pi]$.   

\end{description}

\subsection{Advanced features: map and $\alm$ manipulation}
The package offers several advanced features that allow the users to
manipulate pixelized maps or sets of $\alm$.

\begin{description}
\item{\bf difmap}

{\tt difmap} performs linear combinations of maps, map rotation,
conversion of a map from Galactic to equatorial coordinate base. It
can display simple map information such as pixel numbers and
resolution, histogram, mean and variance, maximum and minimum
temperature values and positions in $(\theta, \phi)$.

\item{\bf difalm} 

{\tt difalm} performs arithmetic operations over $\alm$ such as addition, subtraction, multiplication and division by a factor. It
also allows to swap the amplitudes and phases between two sets of $\alm$. It can multiply/divide $\alm$ by given vectors $v(\l)$. 

\item{\bf mapcut} 

{\tt mapcut} modifies a map in various ways. It can single out an area (either
circular or rectangular) and set temperature values to zero in the
area or outside the area. Furthermore, it can systematically remove or withhold several circular (or rectangular) areas by reading in a file containing those areas' positions and radii (or two corners). It can dissect a rectangular area from a map and output to a file in FITS format.
 
{\tt mapcut} can also display temperature values of a GLESP map at 
given positions in one of the 3 coordinate systems: spherical polar,
Galactic, and equatorial.

\item{\bf mappat} 

{\tt mappat} performs the conversion between ASCII and FITS binary
formats for a GLESP map. The derivative of this is that it allows one to
input any patterns such as point source positions in ASCII and
produce their whole-sky distribution.  
 
\end{description}

\subsection{Conversion between HEALPix and GLESP}
The package also provides map conversion between HEALPix
(\cite{healpix}) and GLESP.
\begin{description}

\item{\bf f2map}

{\tt f2map} performs repixelization of a GLESP map to HEALPix
or to equidistant grid FITS Basic format. In the process of conversion
to HEALPix, {\tt f2map} can perform repixelization either using spline interpolation algorithm or direct mapping. 

\item{\bf cmap}

{\tt cmap} performs repixelization of a HEALPix map to GLESP
using spline interpolation for re-pixelization. On conversion to
GLESP, the resolution is determined by the $N_{side}$ value in
HEALPix, otherwise, the resolution can be set according to user's
input. 

{\tt cmap} 
can also display temperature values of a HEALPix map, or output the
HEALPix map in ASCII format in 3 columns: $\theta$, $\phi$ and
temperature. {\tt cmap} also reads the 2nd field of Binary Table
Extension in HEALPix maps. 
 
\item{\bf ntot}

{\tt ntot} calculates the total pixel number and size for
both GLESP and HEALPix schemes. By giving a resolution, {\tt ntot}
calculates the parameters of the nearest allowed resolution for both schemes.

\end{description}

\section{Objectives}
\subsection{Operation on a single map}
\begin{tabbing}
ana\= yse \hspace{6.3cm}\=  ana\kill

Decomposing a map for its $\alm$/$\cl$                     \>\> {\tt cl2map} \\
Simulating a Gaussian map from given $\cl$             \>\> {\tt cl2map} \\
Displaying  the map $T_{max}$/$T_{min}$, map resolution \\
  \hspace{0.2cm}  mean/variance                           \>\> {\tt difmap} \\
Map rotation                                           \>\> {\tt difmap} \\
Arithmetic operations                                 \>\> {\tt difmap} \\
Producing the histogram                                  \>\> {\tt difmap} \\
Converting the map coordinate system to  \\
  \hspace{0.2cm} equatorial coordinates                 \>\> {\tt difmap} \\
Casting a mask on a map                                \>\> {\tt difmap} \\
Plotting the map and putting markers on the  \\
  \hspace{0.2cm}  figure                           \>\> {\tt f2fig } \\
Displaying the temperature value at a given  \\
  \hspace{0.2cm} coordinate                      \>\> {\tt mapcut} \\
Truncating the temperature fluctuation to         \\
  \hspace{0.2cm} a given temperature range            \>\> {\tt mapcut} \\
Removing or keeping specified areas from      \\
  \hspace{0.2cm} the map                      \>\> {\tt mapcut} \\
Map convolution and deconvolution                      \>\> {\tt rsalm } \\

\end{tabbing}

\subsection{operation on a single set of $\alm$}
\begin{tabbing}
ana\= yse \hspace{6.3cm}\=  ana\kill
Calculating the amplitude $|\alm|$       \>\> {\tt alm2dl} \\
Calculating the phases $\phi_{\lm}$      \>\> {\tt alm2dl} \\
Synthesizing $\alm$ for the map        \>\> {\tt cl2map} \\
Arithmetic operations on $\alm$      \>\> {\tt difalm} \\
Shifting phases                         \>\> {\tt difalm} \\
\end{tabbing}

\subsection{Operations over several maps}
\begin{tabbing}
ana\= yse \hspace{6.3cm}\=  ana\kill
Linear combinations over maps  \>\> {\tt difmap} \\
Cross correlation of 2 maps    \>\> {\tt difmap} \\
Multiplication of 2 maps       \>\> {\tt difmap} \\

\end{tabbing}

\subsection{Operation over several sets of $\alm$}
\begin{tabbing}
ana\= yse \hspace{6.3cm}\=  ana\kill
Linear combinations of $\alm$               \>\> {\tt difalm} \\
Swapping phases and amplitudes of 2 $\alm$  \>\> {\tt difalm} \\
\end{tabbing}

\subsection{Conversions}
\begin{tabbing}
ana\= yse \hspace{6.3cm}\=  ana\kill
Converting GLESP FITS to HEALPix FITS   \>\> {\tt cmap  } \\
Converting HEALPix FITS to GLESP FITS   \>\> {\tt f2map} \\
Converting FITS binary to ASCII         \>\> {\tt mappat} \\
Converting/creating FITS binary from ASCII \>\> {\tt mappat} \\
Information on pixel number and size     \>\> {\tt ntot} \\
\end{tabbing}

\section{Acknowledgment of using GLESP software package}
Those who use the GLESP package for producing results should
acknowledge in the publication. The reference should include the
paper introducing the GLESP concept : Doroshkevich et
al. International Journal of Modern Physics D, Vol 14, No.*,
*** (2005).

\section{Problems and Suggestions}
For any technical issues, bugs, problems or suggestions, please send email
to {\tt glesp@nbi.dk}.


\begin{thebibliography}{}
\bibitem[Doroshkevich et al. 2005]{glesp}
  Doroshkevich, A. G. et al. 2005, Int. J. Mod. Phys. D, Vol. 14,
  No.*, *** 

\bibitem[G\'orski, Hivon \& Wandelt 1999]{healpix}
  G\'orski, K. M., Hivon, E., \& Wandelt, B. D.,
        in ``Evolution of Large--Scale
        Structure: from Recombination to Garching'', 1999
         (http://www.eso.org/science/healpix).




\end{thebibliography}
\end{document}